\documentstyle[aps,prd,preprint,eqsecnum]{revtex}
\begin{document}
\draft
\title{Current commutator anomalies in finite-element quantum electrodynamics}
\author{Dean F. Miller\thanks{E-mail: dfmiller@phyast.nhn.uoknor.edu}}
\address{Department of Physics and Astronomy, University of Oklahoma, Norman,
Oklahoma 73019}
\maketitle
\begin{abstract}
Four-dimensional quantum electrodynamics has been formulated on a
hypercubic Minkowski finite-element lattice.  The equations of motion
have been derived so as to preserve lattice gauge invariance and have been
shown to be unitary.  In addition,
species doubling is avoided due to the nonlocality of the
interactions.  The model is used to investigate the lattice current
algebra.  Regularization of the current is shown to arise in a natural
and nonarbitrary way.  The commutators of the lattice current are
calculated and shown to have the expected qualitative behavior. These
lattice results are compared to various continuum calculations.
\end{abstract}
\pacs{11.15.Ha,11.15.Tk,12.20.Ds,11.40-q}

\narrowtext
\section{INTRODUCTION}
\label{sec:intro}

The linear finite-element approach has previously been successfully applied to
quantum electrodynamics \cite{bender85,miller92}.  The equations of
motion comprise a self-consistent lattice gauge theory, that is, they are
gauge covariant to all orders of the lattice spacing.  Also, the lattice Dirac
equation has been shown to be unitary, {\em i.e.} the fermion
canonical anticommutation relations are preserved in time.  This is
true even in the presence of background electromagnetic fields. The
lattice Maxwell's equations are shown to preserve unitarity in
the absence of interactions.  The interacting case is less clear due
to anomalous behavior which this article will begin to address. In
addition, it has been shown that this model avoids the common lattice
problem of fermion species doubling \cite{bender83}.  The
no-go theorem \cite{nogo} is avoided by the nonlocality of the
interaction term in the Dirac equation.

The finite-element approach has been successfully applied to massless
two-dimensional quantum electrodynamics (the Schwinger model)
\cite{bender85,grose88}.
The axial-vector divergence anomaly was calculated to be
\begin{equation}
\langle\partial_\mu {j_5^\mu}\rangle=-{e^2\over
M\sin(\pi/M)} E.
\end{equation}
Here, $j_5^\mu$ is the axial-vector current, $e$ is the electric
charge, $M$ is the number of spatial lattice sites, and $E$ is the
electric field strength.  The derivative is taken according to the
finite-element prescription: a forward difference in the direction of
the derivative, and a forward average in the other directions. The
relative error is of order $M^{-2}$ as is expected from a linear
finite-element approach.  The anomaly in this model has also recently
been evaluated by a lattice loop calculation of the vacuum
polarization\cite{miller93}

In this paper, we calculate the
commutators of the vector current in four dimensions.  These are
shown to be roughly consistent with results obtained in the continuum.  In
addition, the regularization of the fermion bilinear ${\overline \psi}
\bbox{\gamma} \psi$ is shown to be a direct result of requiring gauge
invariance and unitarity.

In Sec.~\ref{sec:revcomm} we briefly review several continuum
calculations of these commutators.  The ambiguity inherent in these
results is illustrated.  In Sec.~\ref{sec:eoms} our model is introduced in the
form of the four-dimensional lattice equations of motion.  Several
important features of these are presented.  In Sec.~\ref{sec:vector}
we present results for vector-vector
commutators, and in Sec.~\ref{sec:conclusions} we present our conclusions.

\section{CURRENT COMMUTATOR ANOMALIES IN THE CONTINUUM}
\label{sec:revcomm}

It is well known that there are anomalies in the commutators of the
electromagnetic currents.  Schwinger first calculated a non-zero value
for the commutator of the charge density and vector current using a
point-splitting regularization of the current \cite{schwinger59},
\begin{equation}
j_\mu \left( x \right) \equiv \lim_{\bbox{\epsilon} \to 0} e
\psi^{\dag} \left(x-{1\over 2}\bbox{\epsilon}\right)\gamma^0\gamma_\mu\psi
\left(x+{1\over 2}\bbox{\epsilon}\right), \label{eqn:splitcurrent}
\end{equation}
where $\bbox{\epsilon}$ is space-like and the limit is taken
symmetrically in space. Using this current, he arrives at a value for
the commutator of
\begin{equation}
i \langle \left[ J^0\left(0, {\bf x} \right),{\bf J}
\left(0\right)\right]\rangle=-S \bbox{\nabla} \delta\left({\bf
x}\right),  \label{eqn:comm1}
\end{equation}
where $S$ is the divergent limit of
$(2e^2/3\pi^2)(\bbox{\epsilon}^2)^{-1}$.

Later calculations \cite{bjorken66,johnson66,boulware69,chanowitz70}
revealed an additional finite contribution to this commutator.
Its value depends on the type of regularization chosen to
regulate the bilinear current.  Hence, Eq.~(\ref{eqn:comm1}) is
modified to read
\begin{equation}
i \langle\left[ J^0\left(0,{\bf x}\right),{\bf J}\left(0\right)
\right]\rangle= -S \bbox{\nabla} \delta\left({\bf x}\right) -
{d\over\pi^2} \bbox{\nabla}\nabla^2 \delta\left({\bf x}\right),
\label{eqn:comm2}
\end{equation}
where $S$ is the divergent Schwinger term and $d$ depends on the
method of calculation.

The Bjorken-Johnson-Low method defines the equal-time commutator in
terms of the high-energy limit of the time-ordered product
\begin{eqnarray}
\lim_{p_0 \to \infty} -i p_0 \int &&d^4x e^{i p x} \langle T
A\left(x\right) B\left(0\right)\rangle\nonumber\\ &&\equiv \int d^3 x
e^{-i {\bf p}\cdot{\bf x}}\langle\left[ A\left(0,{\bf x}\right), B
\left(0\right)\right]\rangle. \label{eqn:bjl}
\end{eqnarray}
This is shown \cite{johnson66,chanowitz70} to be equivalent to the following
position-space prescription:
\begin{eqnarray}
\langle\left[ A \right.&&\left.\left(0,{\bf
x}\right),B\left(0\right)\right]\rangle\nonumber\\
&&=\lim_{\eta \to 0^+} \left[\langle A\left(\eta,{\bf x}\right)B\left(0\right)
\rangle - \langle B\left(0\right)A\left(-\eta,{\bf x}\right)
\rangle\right].
\end{eqnarray}
Using this prescription, the value of $d$ in Eq.~(\ref{eqn:comm2}) is
calculated to be $d=1/12$ \cite{boulware69,chanowitz70}.

The value of the finite piece of Eq.~(\ref{eqn:comm2}) can also be
calculated using a point-split current such as defined in
Eq.~(\ref{eqn:splitcurrent}). Boulware and Jackiw use this to derive a
value of $d=1/60$ \cite{boulware69}. These authors also use a
generalized point-split current, allowing $\epsilon$ to be a general
four-vector, to derive a value of $d=1/96$. To further illustrate the
ambiguity in this calculation, we give the results of Chanowitz
\cite{chanowitz70}. He shows that is one uses an unsymmetrical
definition of the current
\begin{equation}
J^\mu \left(x\right)=\lim_{\epsilon \to0} \psi^{\dag}
\left(x+\bbox{\epsilon} \right)\gamma^0 \gamma^\mu \psi\left(x\right),
\end{equation}
with $\bbox{\epsilon}$ spacelike, then a value of $d=1/15$ is derived.

Evidently, the results for Eq.~(\ref{eqn:comm2}) depend on the {\it
definition} one chooses for the regulation of the current.  Of course,
this is to be expected, since the quantity is divergent.  Presumably,
the exact value of the coefficients of the Schwinger terms will have
no physical consequence.  However, their very existence proves to
be problematic in establishing the formal properties of the theory.

\section{The lattice equations of motion and regularization of the
electromagnetic current}
\label{sec:eoms}
\subsection{Lattice Dirac equation}

The finite-element formulation of the gauge-invariant Dirac equation
was carried out in Refs.\ \cite{bender85,miller92}. We will merely quote the
results here, expressed in terms of the spatially averaged fermionic field:
\widetext
\begin{eqnarray}
{i \gamma^0 \over \Delta} \left(\psi_{{\overline {\bf m}},n+1} -
\psi_{{\overline{\bf m}},n} \right) & + & {2i \gamma^j
\over\Delta}\left[ \left(\sum_{m_j^\prime=1}^{m_j-1}-
\sum_{m_j^\prime  = m_j+1}^M \right) \left(-1\right)^ {m_j
+m_j^\prime}\psi_{{\overline m}_j^\prime,{\overline{\bf
m}}_\perp,{\overline n}}  \right]\nonumber\\ & + &\mbox{\hspace{0.05in}}
\mu\psi_{{\overline{\bf m}},{\overline n}} -
{2i\gamma^0\over\Delta}\sum_{n^\prime=0}^n {\cal I}^0_
{{\bf m},n,n^\prime}\psi_{{\overline {\bf m}},n^\prime}
-{2i\gamma^j\over\Delta}\sum_{m_j^\prime=1}^M  {\cal I}^j_{{\bf
m}_\perp;m_j m_j^\prime,n}\psi_{{\overline m}_j^\prime, {\overline{\bf
m}}_\perp,{\overline n}}=0, \label{eqn:dirac}
\end{eqnarray}
\narrowtext
\noindent where a sum over the repeated index $j$ is understood, and
the overbar represents a forward average over that coordinate:
\begin{equation}
x_{\overline m}\equiv {1\over2}\left(x_{m+1}+x_m\right).
\end{equation}
(Recall \cite{bender85} that with M odd, $\psi$ is periodic on the
spatial lattice.) We have chosen a hypercubic lattice with lattice
spacing $\Delta$.

Eq.~(\ref{eqn:dirac}) was explicitly constructed to be invariant
under the local gauge transformation
\begin{eqnarray}
\psi_{{\overline {\bf m}},{\overline n}}& \to &
e^{ie\Lambda_{{\overline {\bf m}},{\overline n}}}
\psi_{{\overline {\bf m}},{\overline n}},\nonumber\\ & & \nonumber\\
A_{{\bf m},n}^0 & \to &
A_{{\bf m},n}^0 + {1\over\Delta}\left(\Lambda_{{\overline {\bf m}},n+1} -
\Lambda_{{\overline {\bf m}},n}\right),\nonumber\\ & & \nonumber\\
A_{{\bf m},n}^j & \to &
A_{{\bf m},n}^j + {1\over\Delta}\left(\Lambda_{m_j+1,{\overline {\bf
m}_\perp},{\overline n}} -
\Lambda_{m_j,{\overline {\bf m}_\perp},{\overline n}}\right).
\label{eqn:gaugetrans}
\end{eqnarray}
This particular choice of transformation is made so that the mass term
in the Dirac equation (\ref{eqn:dirac}) is automatically covariant.

To simplify things, we choose the temporal gauge ($A^0=0$).  In this
gauge, ${\cal I}^0$ vanishes identically\cite{bender85}, and
Eq.~(\ref{eqn:dirac})
contains only fields at time {\it n}.  The explicit form of the
spatial parts of the interaction is
\widetext
\begin{eqnarray}
{\cal I}_{{\bf m}_\perp;m_j,m_j^\prime,n}^j & = &
\epsilon_{m_j,m_j^\prime} \left(-1\right)^{m_j +m_j^\prime}\left[
-1 +\cos\left( \sum_{m_j^{\prime\prime}=1}^M {\text{sgn}}\left(m_j^{
\prime\prime}-m_j\right)  {\text{sgn}}\left(m_j^{\prime\prime}-
m_j^\prime\right) \zeta^j_{{\bf m}_\perp,m_j^{\prime\prime},n}\right)\sec
\zeta^j_{{\bf m}_\perp,n} \right]  \nonumber \\
  & + & \mbox{\hspace{0.05in}}i\left(-1\right)^{m_j+m_j^\prime}
\sin\left(\sum_{m_j^{\prime\prime}= 1}^M
{\text{sgn}} \left(m_j^{\prime\prime}- m_j\right){\text{sgn}}\left(
m_j^{\prime\prime}-  m_j^\prime\right)\zeta^j_{{\bf m}_\perp,
m_j^{\prime\prime},n}\right)\sec\zeta^j_{{\bf m}_\perp,n}.
\label{eqn:interaction}
\end{eqnarray}
\narrowtext
\noindent We have used the abbreviations
\begin{equation}
\zeta^j_{{\bf m}_\perp,m_j,n}={e\Delta\over 2}A_{{\bf m}_\perp, {\overline
{m_j -1}},n}^j, \;\;\;\zeta^j_{{\bf m}_\perp,n}=\sum_{m_j=1}^M \zeta^j_{{\bf
m}_\perp,m_j,n},
\end{equation}
and
\begin{eqnarray}
\text{sgn}\left(x\right)=&&\left\lbrace
\begin{array}{rl}
+1,\; & x>0\\
-1,\; & x\leq0
\end{array}\right.\\ & & \nonumber\\
\epsilon_{m_j,m_j^\prime}=&&\left\lbrace
\begin{array}{rl}
+1,\; & m_j>m_j^\prime\\
0,\; & m_j=m_j^\prime\\
-1,\; & m_j<m_j^\prime.
\end{array}\right.
\end{eqnarray}

Eq.~(\ref{eqn:dirac}) can be written in a more compact matrix
notation:
\begin{equation}
{i\over\Delta}\left(\phi_{n+1} -\phi_n\right) + {2 i \gamma^0
\bbox{\gamma}\over\Delta}\cdot\left({\bf Q}-{\bf I}_n\right)
\phi_{\overline n} +\mu\phi_{\overline n} = 0.
\label{eqn:matdirac}
\end{equation}
Here, ${\bf Q}$ and {\bf I} are matrices in the spatial
indices
\begin{eqnarray}
\left({\bf Q}^j\right)_{{\bf m},{\bf m}^\prime} & =
& \delta_{{\bf m}_\perp,{\bf m}_\perp^\prime}\left(-1\right)
^{m_j + m_j^\prime}\epsilon_{m_j,m_j^\prime},
\nonumber\\ & & \nonumber\\ & & \nonumber\\
\left({\bf I}_n^j\right)_{{\bf m},{\bf m}^\prime} & =
& \delta_{{\bf m}_\perp,{\bf m}_\perp^\prime}
{\cal I}_{{\bf m}_\perp;m_j,m_j^\prime,n}^j,
\label{eqn:d&Imatrices}
\end{eqnarray}
and $\phi_n$ is a vector of the spatially-averaged fields
\begin{equation}
\left(\phi_n\right)_{\bf m}=\psi_{{\overline {\bf m}},n}.
\end{equation}
Notice that {\bf Q} is related to the lattice derivative,
\begin{equation}
{2\over\Delta}{\bf Q} \leftrightarrow \bbox{\nabla},
\end{equation}
and the lattice version of the covariant derivative is
\begin{equation}
{2\over\Delta}\left({\bf Q}-{\bf I}_n\right) \leftrightarrow
\bbox{{\cal D}}.
\end{equation}

Eq.~(\ref{eqn:matdirac}) can be solved for the transfer matrix defined
by
\begin{equation}
\phi_{n+1}\equiv \text{T}_n\phi_n,
\label{eqn:Tdef}
\end{equation}
giving
\begin{equation}
\text{T}_n = {1 - \gamma^0\bbox{\gamma}\cdot\left({\bf Q} -
{\bf I}_n\right) + i \nu\gamma^0 \over{1 + \gamma^0\bbox{\gamma}
\cdot\left({\bf Q} -{\bf I}_n\right) - i \nu\gamma^0}},
\label{eqn:transfer}
\end{equation}
with $\nu ={\mu\Delta\over2}$. From Eqs.~(\ref{eqn:interaction})
and (\ref{eqn:d&Imatrices}), it is clear that both ${\bf Q}$
and ${\bf I}_n$ are anti-hermitian.  Therefore, the transfer
matrix defined in Eq.~(\ref{eqn:transfer}) is unitary. Hence, the
spatially averaged fields $\phi_n$ can be taken to be the canonical
fermionic fields.

\subsection{Lattice Maxwell equations}

Using the notation of Eq.~(\ref{eqn:d&Imatrices}), the lattice
Maxwell's equations are written as
\begin{eqnarray}
{\bf E}_{\overline{n}} & = & {1\over\Delta}\left(
{\bf A}_{n+1}-{\bf A}_{n}\right),\nonumber\\
{2\over\Delta}{\bf Q}\cdot{\bf E}_{\overline{n}} & = & J^0_n, \nonumber\\
{\bf E}_{\overline{n+1}}-{\bf E}_{\overline{n}} & = &
-{\bf J}_n - {2\over\Delta}{\bf Q}\cdot\bbox{\cal F}_{\overline{n}},\nonumber\\
{\cal F}^{i j}_{\overline{n}} & = &
-{2\over\Delta}\left({\bf Q}^i A^j_{\overline{n}}
-{\bf Q}^j A^i_{\overline{n}}\right),
\label{eqn:maxwells}
\end{eqnarray}
\noindent where
\begin{eqnarray}
{\bf E}_n & \equiv & {\bf E}_{\overline{\bf m},n},\nonumber\\
{\bf A}_n & \equiv & {\bf A}_{\overline{\bf m},n},
\end{eqnarray}
\noindent Gauge invariance of Eqs. (\ref{eqn:maxwells}) under
(\ref{eqn:gaugetrans}) is assured as
long as the current $J^\nu_{{\bf m},n}$ is constructed to be gauge invariant.

\subsection{Current regularization}

The simplest choice of gauge-invariant current is
\begin{eqnarray}
J_{{\bf m},n}^\mu & \equiv & e{\overline \psi}_{{\overline{\bf
m}},{\overline n}}
\gamma^\mu {\psi_{{\overline{\bf m}},{\overline n}}} \\
& = & e{\overline \phi}_{{\bf m},{\overline n}}\gamma^\mu
\phi_{{\bf m},{\overline n}},
\label{eqn:current}
\end{eqnarray}
\noindent which is invariant under the transformation
(\ref{eqn:gaugetrans}).
Notice that the covariant field $\psi_{{\overline{\bf m}},{\overline
n}}$ that is
involved in the definition of
$J_{{\bf m},n}^\mu$ is not the same as the canonical field $\phi_{{\bf
m},n}$ defined in
Eq.~(\ref{eqn:Tdef})---they differ by a temporal averaging. This forces
point-splitting of the current:
\widetext
\begin{equation}
J_{{\bf m},n}^\mu=e{\overline \phi}_{{\bf m},{\overline n}}\gamma^\mu
\phi_{{\bf m},{\overline n}}
=e\sum_{{\bf m}^\prime,{\bf m}^{\prime\prime}}\phi_{{\bf
m}^\prime,n}^{\dag}
\left({{1+\text{T}^{\dag}\left(A^\mu_n\right)}\over2}\right)
_{{\bf m}^\prime,{\bf m}}\gamma^0\gamma^\mu
\left({1+\text{T}\left(A^\mu_n\right)\over2}\right)_{{\bf m}, {\bf
m}^{\prime\prime}}\phi_{{\bf m}^{\prime\prime},n}.
\label{eqn:current2}
\end{equation}
\narrowtext
\noindent The important point is that this is not introduced in an arbitrary
way.   {\it The regularization of the
current is mandated by requiring gauge-invariance and unitarity of the
equations of motion.}

\section{Vector-vector commutators on the lattice}
\label{sec:vector}
\subsection{Analytical results}

Using the definition of the electromagnetic current
Eq.~(\ref{eqn:current}), the commutators of various components can be
calculated.  To lowest order in $e$, the commutators can be calculated
using the zeroth-order contribution of the transfer matrix
\begin{equation}
\text{T}_n = {{1 - \gamma^0\left(\bbox{\gamma}\cdot{\bf Q}\right)
+i\nu\gamma^0}\over{ 1 + \gamma^0\left(\bbox{\gamma}\cdot
{\bf Q}\right) -i\nu\gamma^0}}
\label{eqn:Tfree}
\end{equation}
and a free field Fock space expansion for the Dirac fields
\begin{eqnarray}
\phi_{{\bf m},n}=\sum_{\sigma,{\bf p}}&&
\sqrt{{\mu\over\omega_{\bf p}}}\left[u_{\bf p}^{\left(\sigma\right)}
b_{\bf p}^{\left(\sigma\right)}e^{2\pi i{\bf p}\cdot{\bf m}/M}
\right.\nonumber\\
&&\left.\mbox{\hspace{0.5in}}+ v_{\bf p}^{\left(\sigma\right)}d_{\bf
p}^{\left(\sigma\right)\dag}
e^{-2\pi i{\bf p}\cdot{\bf m}/M}\right],
\end{eqnarray}
where the spinors are normalized according to
\begin{eqnarray}
\sum_\sigma u^{\left(\sigma\right)}u^{\left(\sigma\right)\dag}\gamma^0
&=& {{\omega\gamma^0-{2\over\Delta}\left(\bbox{\gamma}\cdot{\bf t}\right)-\mu}
\over{2\mu}}\nonumber\\ & & \nonumber\\
\sum_\sigma v^{\left(\sigma\right)}v^{\left(\sigma\right)\dag}\gamma^0
&=& {{\omega\gamma^0-{2\over\Delta}\left(\bbox{\gamma}\cdot{\bf t}\right)+\mu}
\over{2\mu}}\\ & & \nonumber
\end{eqnarray}
with
\begin{equation}
{\bf t}_{\bf p}^i = \tan\left({p_i\pi\over M}\right),\;\;
\omega_{\bf p}=\sqrt{\left({2\over\Delta}\right)^2{\bf t}_{\bf p}^2 + \mu^2}.
\end{equation}
Then the canonical anticommutation relations for the Dirac fields
\begin{equation}
\left\{\phi_{{\bf m},n},\phi_{{\bf m}^\prime,n}^{\dag}\right\rbrace =
{1\over\Delta^3}\delta_{{\bf m},{\bf m}^\prime}
\end{equation}
are satisfied if
\begin{eqnarray}
\left\{b_{\bf p}^\sigma,b_{{\bf p}^\prime}
^{\sigma^\prime\dag}\right\} &=&
{1\over\left(M\Delta\right)^3}\delta_{{\bf p},{\bf
p}^\prime}\delta^{\sigma,\sigma^\prime}\nonumber\\
& & \nonumber\\
\left\{d_{\bf p}^\sigma,d_{{\bf p}^\prime}
^{\sigma^\prime\dag}\right\} &=& {1\over\left(M\Delta\right)^3}
\delta_{{\bf p},{\bf p}^\prime}\delta^{\sigma,\sigma^\prime},
\end{eqnarray}
and all other anticommutators of these operators vanish.

The results of the calculations are presented below:
\widetext
\begin{eqnarray}
\left\langle\left[J_{{\bf m},n}^0,J_{{\bf m}^\prime,n}^0\right]
\right\rangle&=&0 \\ & &\nonumber\\
\left\langle\left[J_{{\bf m},n}^i,J_{{\bf m}^\prime,n}^j\right]
\right\rangle&=&0 \\ & &\nonumber\\
i\left\langle\left[J_{{\bf m},n}^0,J_{{\bf m}^\prime,n}^j\right]
\right\rangle &=& -{4 i e^2 \over \left(M\Delta\right)^6}
\sum_{{\bf p},{\bf p}^\prime} {\tan\left({\pi{\bf
p}_j^\prime\over2}\right)\over\left(1+\left({\Delta\omega_{{\bf p}- {\bf
p}^\prime}\over2}\right)^2 \right)\left(1+ \left({\Delta\omega_{{\bf
p}^\prime}\over2}\right)^2\right)
\left({\Delta\omega_{{\bf p}^\prime}\over2}\right)}e^{2\pi i {\bf p}
\cdot\left( {\bf m} -{\bf m}^\prime \right)/M}.
\label{eqn:commval}
\end{eqnarray}
\narrowtext

\subsection{Numerical evaluation}

The matrix (in {\bf m} and ${\bf m}^\prime$) represented by
Eq.~(\ref{eqn:commval}) is shown for lattice size
$M=9$ and mass $\mu=0$ in Fig.~\ref{fig:3dplot}. The abscissas of the
plot correspond
to one-dimensional representations of the three components of {\bf m}
and ${\bf m}^\prime$, respectively.  (That is, the base-$M$ number
($m_1,m_2, m_3$) is converted to a base-10 value of the abscissa.) It
is apparent that the leading-order behavior of this commutator is a
first derivative of the Dirac delta function, as expected.

To make this comparison more quantitative, we fit this result to the
functional form expected in the continuum, Eq.~(\ref{eqn:comm2}).
The lattice analog of the Dirac delta function is
\begin{equation}
\delta\left({\bf x}\right) \approx {1\over\Delta^3}\delta_{{\bf
m},{\bf 0}}={1\over\left(M\Delta\right)^3}\sum_{\bf p} e^{2\pi i {\bf
p} \cdot {\bf m}/M},
\label{eqn:delta}
\end{equation}
so we take as a trial function
\widetext
\begin{equation}
i\left\langle\left[J_{{\bf m},n}^0,J_{{\bf m}^\prime,n}^j\right]
\right\rangle \approx
{-e^2\over\left(M\Delta\right)^3}{2\pi i\over M\Delta}\sum_{\bf p} p_j
\sum_{r=0}^{R-1} a_{2r+1}P_r\left(\Delta\right)(-1)^r \left({2\pi \over
M\Delta} \right)^{2r} p^{2r} e^{2\pi i {\bf p}\cdot \left({\bf m} -
{\bf m}^\prime\right)/M}.
\label{eqn:trialcomm}
\end{equation}
\narrowtext
\noindent The coefficients $a$ do not depend on the lattice spacing
$\Delta$ (for mass $\mu = 0$); they presumably remain finite in the
continuum limit,
$\Delta \to 0$. The continuum behavior of the terms in
(\ref{eqn:trialcomm}) is dictated by the functions
$P_r\left(\Delta\right)$.  These are inserted to
force each term in the series to be of the same order in $\Delta$ as
the commutator (\ref{eqn:commval}), namely $1/\Delta^6$:
\begin{equation}
P_r\left(\Delta\right) = \Delta^{2r-2}.
\end{equation}
Therefore, in the continuum limit, the first term in
Eq.~(\ref{eqn:trialcomm}) will be quadratically divergent, the second will
be finite, and the rest will vanish, as expected.

To perform the fit, the data is first converted from a
three-dimensional to a two-dimensional
representation. (This is necessary because the generation of 3-D plots
is impractical beyond $M\simeq9$ and because 3-D fits are prone to
difficulties.) Specifically, a plot such as Fig.~\ref{fig:3dplot} is
projected onto a plane orthogonal to the $m,m^\prime$ plane and to the
diagonal of the matrix.  This yields a graph such as the solid curve in
 Fig.~\ref{fig:2dproject}(a). The curve is then fit to a similar
projection of the trial function in Eq.~(\ref{eqn:trialcomm}).  In
particular, Fig.~\ref{fig:2dproject} shows {\em fit vs.\ data} for a
fit of the commutator (\ref{eqn:commval}) to
a linear superposition of three spectra, {\em i.e.}, $R=3$ in
(\ref{eqn:trialcomm}).

It turns out that a fit to five spectra
produces comparable results. The addition of the last two spectra
neither particularly improves the fit nor radically changes the
coefficients of the original three.  (That is, the fit is stable.) In
addition, the
coefficients of all but the first two spectra are vanishingly small. In
Fig.~\ref{fig:posresults} we show the results for a fit to
(\ref{eqn:trialcomm}) with $R=3,4, \text{and}\; 5$ for various lattice
sizes.  These are compared to some of the results of continuum
calculations.  The lattice values, though of the same order as other
results, are significantly different.  This is as expected, since
(\ref{eqn:current2}) is yet another definition of current
regularization. (Note that the result of Chanowitz
\cite{chanowitz70} is within one sigma of our result.  This is
interesting in that both his method and ours use a temporal averaging
to regularize the current.)

A note should be added to discuss the parameterization of errors in the
data:  Since the data (\ref{eqn:commval}) is computer-generated, any
``measurement'' errors should be strictly due to round-off errors.
Presumably this will be roughly the same for each data point.
Therefore, it is reasonable to assume that all measurement errors are
equal, $\sigma_i^2~\equiv~\sigma^2$. If we further assume normally
distributed errors, then the value of $\chi^2$ for the fit is
\begin{equation}
\chi^2 = {1\over\sigma^2}\sum_{\text{data points}}\left(\text{data -
fit}\right)^2.
\end{equation}
If we now require the fit to be ``good'', {\em i.e} $\chi^2$ per
degree of freedom equal to one, then we can determine a value for
$\sigma^2$,
\begin{equation}
\sigma^2 = {\sum_{\text{data points}}\left(\text{data - fit}\right)^2\over
\text{degrees of freedom}},
\end{equation}
where the number of degrees of freedom is the number of data points
less the number of terms in the trial function.
Multiplying this by the diagonal elements of the covariance matrix
$C$ (which is determined as a by-product of our minimization process),
gives the variances of the coefficients $a_r$
\begin{equation}
\sigma^2\left(a_r^j\right)=\sigma^2 C_{jj}.
\end{equation}
These are plotted along with the values of the coefficients in
Fig.~\ref{fig:posresults}.

There are a few features of the graphs of Fig.~\ref{fig:posresults}
which
merit some discussion.  First, there is the consistent kink in the
data at lattice size $M=13$. We have no explanation for this behavior
except to suppose it is just a numerical quirk of our code.

Second, the scattering of the data points increases significantly with
the number of fit spectra $R$. This seems to be a problem with
numerical precision in the fitting procedure---Each successive term in
the trial function (\ref{eqn:trialcomm}) is an order of magnitude or two
larger than the previous one.  Thus, at higher numbers of fit spectra
($R\sim 5$), the range in values of terms in (\ref{eqn:trialcomm})
approaches the numerical precision of the data.  For example, if we
continue to $R=6$, our procedure breaks down---the coefficients
generated by the fitting algorithm do not give a good fit to the data.

Lastly, the errors associated with the data
do not seem to account for the scatter of the points (especially for
$R=5$.)  Remember, though, that the derivation used to give the errors
in our data is based upon several assumptions. These
are not rigorously justified.  Rather these suppositions serve to
compensate for our lack of knowledge of the fundamental source of
errors.  In order to summarize the data, incorporating the
errors in a more model-independent manner, we have averaged the data
points at all lattice sizes (excluding $M=13$) and plotted these
results alongside the other data. (Figure~\ref{fig:posresults})

The data for $R=3$ and $R=4$ are in good agreement over a
wide range of lattice sizes.  These only disagree at low $M$, where
the model is not expected to yield good results, and at $M\sim 31$,
where the fitting procedure for $R=4$ begins to break down.  This
agreement gives support to our claim that the fit is stable with
respect to the number of fit spectra $R$.  Thus, we are able to
consider the best data (with $R=3$) as representative of the
commutator (\ref{eqn:commval}).  The results in the fit of the trial
function (\ref{eqn:trialcomm}) with $R=3$ to the commutator
(\ref{eqn:commval}) are given in Fig.~\ref{fig:3results}.

\section{Conclusions}
\label{sec:conclusions}

Using the definition of the current (\ref{eqn:current2}), the current
commutators are calculated and are seen to be anomalous.  These
exhibit the qualitative behavior
expected from continuum calculations, having both quadratically
divergent and finite contributions. The coefficients of
these are of the same order of magnitude as the various continuum
calculations. In fact, our results seem to be in quantitative
agreement with those of Chanowitz \cite{chanowitz70} who also uses a
temporal point-splitting regularization scheme.

Furthermore, the results arise from a regularization of
the current that is both natural and nonarbitrary. The requirements of
gauge invariance and unitarity dictate a current which exhibits a
temporal averaging of the composite fermionic fields.  This averaging
is distinct, yet akin, to a point-splitting regularization.

We have also begun work on vector-axial-vector commutators.  These are
more interesting since they can be directly related to the chiral
anomaly\cite{hosono88}
\begin{equation}
A\left(x\right) = {1\over4\pi^2}\epsilon^{0ijk}\lbrace
E_i,F_{jk}\rbrace,
\end{equation}
by the Gauss' Law constraint,
\begin{equation}
\bbox{\nabla}\cdot{\bf E} = J^0.
\end{equation}
Preliminary results are promising, but much work remains to be done.

\acknowledgements

I wish to express my gratitude to K.~A.~Milton, S.~Siegemund-Broka,
and P.~Jain for {\em many} helpful discussions. I also want to
thank the U.~S.~Department of Education and the U.~S.~Department
of Energy for financial support.

\begin{figure}
\caption{Three-dimensional plot of (\protect\ref{eqn:commval}) for $M=9$ and
$\mu=0$.}
\label{fig:3dplot}
\end{figure}

\begin{figure}
\caption{(a) Fit {\em vs.\ }data of two-dimensional projection of
(\protect\ref{eqn:commval}) and (\protect\ref{eqn:trialcomm}) with $R=5$.
($M=21$ and $\mu=0$) (b) Smaller scale picture of the same. (The
position variable has been normalized to range from $-1$ to $+1$.)}
\label{fig:2dproject}
\end{figure}

\begin{figure}
\caption{Coefficients of a 3, 4, and 5 spectra fit for different lattice
sizes. $(\mu=0)$ (a) Coefficient of first derivative of delta
function. (b) Coefficient of third derivative of delta functions. The
right-most data points summarize the data excluding $M=13$.  Also
plotted are various continuum results for $d$ in
Eq.~(\protect\ref{eqn:comm2}).}
\label{fig:posresults}
\end{figure}

\begin{figure}
\caption{Summary of the results for a three spectra fit.}
\label{fig:3results}
\end{figure}

\end{document}